\documentstyle [12pt]{article}

\def\tr{{\rm Tr}}
\newcommand{\be}{\begin{eqnarray}}
\newcommand{\ee}{\end{eqnarray}}
\newcommand{\BE} {\begin{equation}}
\newcommand{\EE} {\end{equation}}

\begin{document}

\thispagestyle{empty}
\hspace{4.4in}
OSU-NT-\#93-0128
\vspace{.3in}
\setlength{\unitlength}{1in}
\baselineskip 0.20in
\begin{center}
{\Large\bf Residual Gauge Fixing \\
in Light-Front QCD}
\end{center}
\begin{center}
\ \\
{\bf Wei-Min Zhang and Avaroth Harindranath} \\
{\normalsize{\em Department of Physics, The Ohio State University \\
Columbus, Ohio 43210, USA}} \\

\vspace{.4in}

{\bf abstract}
\end{center}
\begin{quote}
Understanding the nontrivial features of light-front QCD is a central
goal in current investigations of nonperturbative light-front
field theory.  We find that, with the choice of light-front gauge
with antisymmetric boundary conditions for the field variables, the
residual gauge freedom is fixed and the light-front QCD vacuum
is trivial. The nontrivial structure in light-front QCD is
determined by non-vanishing asymptotic physical (transverse)
gauge fields at longitudinal infinity, which are responsible
for nonzero topological winding number.
\end{quote}

\newpage

Recently, the search for nonperturbative solutions of quantum
chromodynamics (QCD) in light-front coordinates has become a very
active subject in hadronic and nuclear physics {\cite{Ji}}.
In the light-front quantization of QCD, one usually chooses the
light-front gauge, $A_a^+=0$.  As is well known in non-abelian
gauge theory, the advantage of choosing a physical gauge, such
as temporal gauge, axial gauge or light-front gauge, is that
no ghost field is introduced. However, there still exists
residual gauge freedom in a physical gauge, which is manifested
differently in equal-time and light-front quantizations.
Fixing the residual gauge freedom is the first
step in nonperturbative calculations in the canonical
Hamiltonian formalism.

In equal-time quantization, one usually chooses the temporal gauge
$(A_a^0=0)$.  The residual gauge freedom in $A_a^0=0$ gauge is determined
by Gauss's law (the generators of residual gauge transformations
annihilate physical states). However, the implementation of Gauss's
law is complicated, and in practice, it has been solved only on the
lattice {\cite{Kougt}}.  Sometimes one chooses the axial gauge $(A_a^3=0)$
in equal-time canonical quantization.  In axial gauge, the
time-component of gauge potentials can be computed explicitly, and
thus there is apparently no residual gauge freedom.
However, as noticed first by Schwinger {\cite{Schwinger}}, there
indeed exist residual gauge transformations in $A_a^3=0$,
which are generated by the longitudinal color electric fields at
infinity in the $z$-direction.  Despite this insight,
investigations of the residual gauge fixing
in axial gauge have remained obscure {\cite{Axialg}}.
In light-front quantization, it is convenient to choose the
light-front gauge. To the best of our knowledge, the issue of residual
gauge fixing in $A_a^+=0$ gauge has not been explored in light-front
coordinates.{\footnote{One may find some discussion about residual
gauge freedom in $A_a^+=0$ in equal-time quantization
{\cite{Bassetto}}.  In this case, the residual gauge
fixing could, in principle, be determined by Gauss's law, as
in the case of temporal gauge.}}

In the light-front canonical quantization of QCD, the residual gauge
freedom in $A_a^+=0$ is associated with the $x^-$-independent gauge
transformations which operate on the modes with zero light-front
longitudinal momentum, i.e., the $k^+=0$ modes.  By choosing
antisymmetric boundary
conditions for physical field variables at light-front longitudinal
infinity, the $k^+=0$ modes are removed and the residual gauge
freedom is fixed.  However, the antisymmetric boundary conditions
imply that the physical fields do not vanish at longitudinal infinity.
In this letter, we show that the asymptotic gauge fields at longitudinal
infinity give rise to a non-vanishing topological winding number.
Thus, the asymptotic gauge fields induced by the residual gauge
fixing may be the source of nontrivial properties in light-front QCD
{\cite{Ji}}.

{\em 1. Light-front QCD}.
We begin by recalling the essential features of the light-front
formulation of QCD (for more detailed discussions see
refs.{\cite{Casher,zhang2}}).
In light-front coordinates $x^{\pm} = x^0 \pm x^3,~ x_{\bot}^i
= (x^1,x^2)$, with the light-front gauge $A_a^+ \equiv A_a^0 +
A_a^3 =0$, the QCD Hamiltonian can be written as follows:
\BE
	H = \int dx^- d^2 x_{\bot} \left\{ \frac{1}{2} (E_a^{-2}
		+ B_a^{-2}) + \psi_+^{\dagger} \{ \alpha_{\bot} \cdot
		( i \partial_{\bot} + g A_{\bot}) + \beta m \} \psi_-
		\right\} ,
\EE
where $E_a^- = - \frac{1}{2}\partial^+ A_a^- $ and $B_a^- = \partial^1
A_a^2 - \partial^2 A_a^1 + g f^{abc}A_b^1 A_c^2$ are the longitudinal
components of color electric and magnetic fields,
and $\psi_{\pm}= \frac{1}{2}\gamma^0\gamma^{\pm} \psi = \Lambda_{\pm}
\psi $ the light-front quark field variables. In fact, the
light-front QCD is a two-component theory, where all the physical
quantities depend only on the two-component physical (transverse)
gauge fields $(A_a^i,~i=1,2)$ and two-component quark fields
{\cite{zhang1}. The dependent field variables $(A_a^-, \psi_-)$
are determined by the following constraint equations,
\be
	& & \psi_- = \frac{1}{i\partial^+} \{ \alpha_{\bot} \cdot
		( i \partial_{\bot} + g A_{\bot}) + \beta m \} \psi_+ ,  \\
	& & A_a^- = - \frac{2}{\partial^+} E_a^-
		= \frac{1}{(\partial^+)^2} (\partial^i\partial^+ A_a^i
		+ gf^{abc} (A_b^i \partial^+A_c^i + 2 \psi_+^{\dagger}
		T^a \psi_+)) .
\ee

In the above formulation, one has to define the operator
$(\frac{1}{\partial^+})$.  A typical definition in light-front
field theory is
\BE
	\left( \frac{1}{\partial^+} \right)^n f(x^+,x^-,x_{\bot})
		= \frac{1}{4^n}	\int_{-\infty}^{\infty} dx_1^- \cdots
		dx_n^- \varepsilon (x^- - x_1^-) \cdots \varepsilon
		(x_{n-1}^- - x_n^-) f(x^+,x_n^-,x_{\bot}),
\EE
where $\varepsilon(x) = 1, 0, -1$ for $x > 0, = 0 , < 0$. Eq.(4)
requires that all the field variables satisfy antisymmetric boundary
conditions.  Thus, the basic commutation relations in phase space
quantization {\cite{zhang2,Jackiw}} become:
\be
	& & [A_a^i (x) , \partial^+ A_b^j(y) ]_{x^+=y^+} = i \delta_{ab}
		\delta^{ij} \delta^3 (x-y) , \\
	& & [A_a^i (x) , A_b^j(y) ]_{x^+=y^+} = - i \delta_{ab} \delta^{ij}
		\frac{1}{4} \varepsilon(x^- - y^-) \delta^2(x_{\bot}-
		y_{\bot}) , \\
	& & \{\psi_+ (x), \psi_+^{\dagger}(y) \}_{x^+=y^+} =
		\Lambda_+ \delta^3(x-y),
\ee
and all other commutators between the physical degrees of freedom vanish.
A consistent definition for eq.(6) implies that
\BE
	\lim_{x^- \rightarrow \infty} \varepsilon (x^- - y^-)
		\equiv \eta(y^-) = \left\{ \begin{array}{ll}
			0 & y^- \rightarrow \pm \infty \\
			1 & {\rm otherwise} \end{array} \right. .
\EE

As a consequence of eq.(4), the choice of antisymmetric boundary
conditions remove the $k^+=0$ modes {\cite{zhang2}}.
Meanwhile, the light-front quantization of QCD ensures
that the light-front longitudinal momentum of quarks and gluons
must be positive semidefinite ($k^+ \geq 0$) {\cite{Kogut}}.
This property implies that the light-front QCD vacuum, which has
zero total momentum, only contains particles with zero longitudinal
momentum {\cite{Kogut}}.  Thus, with the antisymmetric boundary
condition, the light-front QCD vacuum is {\em trivial},
i.e., it is identical to the light-front bare vacuum. On the other
hand, it is known that QCD in equal-time quantization has a
nontrivial vacuum associated with topological gauge solutions
{\cite{thooft}}.

The question is: where are the nontrivial features hidden in the
above formulation of light-front QCD?
Note that the light-front bare vacuum is not identical to the
equal-time bare vacuum.  The trivial light-front vacuum
originates from the choice of antisymmetric boundary conditions
at light-front longitudinal infinity, and with such boundary
conditions the physical field variables do not vanish on the
boundary surfaces.  Since antisymmetric boundary conditions
for the physical fields imply possible existence of nontrivial
topological soliton solutions {\cite{Cheng}}, we suggest that
{\em the nontrivial QCD structure must be carried purely
by the boundary behavior of gauge fields.}  The antisymmetric
boundary condition for the physical (transverse) gauge
fields fixes the residual gauge freedom, and the boundary
behavior of gauge fields determines the nontrivial topological
properties of QCD in $A_a^+=0$ gauge.

{\em 2. Residual gauge transformations}.
In light-front quantization, the residual gauge
transformations may be generated by operators
\BE
	R_a = - \frac{1}{2} \int_{-\infty}^{\infty} dx^-
		\left\{ 2 \partial^+ \partial^i A_a^i
		+ g(f^{abc} A_b^i \partial^+ A_c^i + 2
		\xi^{\dagger} T^a \xi) \right\} .
\EE
The definition of $R_a$ here is different from the residual
gauge transformations in axial gauge ($A_a^3=0$), where the corresponding
generators are defined as the operators $E_a^3$ at $z=\infty$.  In the
light-front gauge, the boundary operators $E_a^-|_{x^- =\infty}$
do not generate the correct gauge transformations.
The $R_a$ in eq.(9) is different by a factor of 2 from the first
term in $E_a^- |_{x^-=\infty}$ [see eq.(3)].  This difference follows
because the $E_a^i= -\frac{1}{2} \partial^+ A_a^i$ are not
dynamical variables. We shall first show how the $R_a$ generate the
transverse residual gauge transformations in light-front QCD.

The transverse gauge transformation in light-front coordinates can be
generally defined by
\be
	& & \psi_+(x) \longrightarrow \psi_+'(x) = u(x_{\bot})
		\psi_+(x), \\
	& & A^i(x)  \longrightarrow {A^i}'(x) = u(x_{\bot}) A^i(x)
		u^{-1}(x_{\bot}) - \frac{i}{g} (\partial^i
		u(x_{\bot}) ) u^{-1}(x_{\bot})
\ee
where $u(x) = \exp(- i \theta_a(x_{\bot}) T^a )$ are $SU(3)$
gauge group elements.  For the infinitesimal $\theta_a(x_{\bot})$,
the above transformations lead to
\be
	& & \delta \psi_+(x) \equiv \psi_+'(x) - \psi_+(x) = - i
		T^a \theta_a (x_{\bot}) \psi_+ (x) , \\
	& & \delta A_a^i (x) \equiv {A_a^i}' (x)  - A_a^i (x) = f^{abc}
		\theta_b(x_{\bot}) A_c^i (x) - \frac{1}{g} \partial^i
		\theta_a(x_{\bot}) .
\ee

The gauge transformations generated by $R_a$ are defined in quantum theory
such that the quark and gluon field operators and states (wave
functions) transform as follows {\cite{Susskind}},
\be
	& & \psi_+(x) \longrightarrow \psi'_+(x) \equiv U
		\psi_+ (x) U^{-1} , \\
	& & A_a^i (x) \longrightarrow A_a^{'i} (x) \equiv
		U A_a^i (x) U^{-1}  \\
	& & | \Phi \rangle \longrightarrow | \Phi \rangle' \equiv
		U | \Phi \rangle ,
\ee
where
\BE
	U = \exp \left\{- \frac{i}{g} \int d^2x_{\bot} \theta_a(x_{\bot})
		R_a (x_{\bot}) \right\}
\EE
{\em and the $\theta_a(x_{\bot})$ vanish at transverse spatial infinity.}
For infinitesimal $\theta_a(x_{\bot})$, eqs.(14--15) are reduced to
\BE
	\delta \psi_+ = i[G_{\theta}~,~ \psi_+] ~~, ~~
		\delta A_a^i = i[ G_{\theta}~ , ~A_a^i],
\EE
where
\BE
	G_{\theta} = - \frac{1}{g} \int d^2x_{\bot} \theta_a(x_{\bot})
		R_a (x_{\bot}).
\EE
Using eqs.(5-7){\footnote{If the commutation involves $\partial^+
A_a^i$, one must use eq.(5) rather than (6). Otherwise, the
ordering of differentiation and integration may cause a
problem.}}, we have
\be
	& & [R_a(x_{\bot}), \psi_+(y^-,y_{\bot})]_{x^+=y^+} =
		g \delta^2(x_{\bot} - y_{\bot}) T^a \psi_+(y^-,
		y_{\bot}) \\
	& & [R_a(x_{\bot}), A_b^i(y^-,y_{\bot})]_{x^+=y^+} = -i g f^{abc}
		\delta^2(x_{\bot} - y_{\bot}) A_c^i (y^-,y_{\bot})
		\nonumber \\
	& & ~~~~~~~~~~~~~~~~~~~~~~~~~~~~~~~~~~~~~~~~~~~~
		+ i \delta_{ab} \partial_x^i \delta^2 (x_{\bot}
		- y_{\bot}) \\
	& & [R_a(x_{\bot}), R_b(y_{\bot})]_{x^+=y^+} = i g f^{abc}
		\delta^2 (x_{\bot} - y_{\bot}) R_c(x_{\bot}).
\ee
{}From eqs.(20--22), it is easy to verify that eq.(18) produces
the local gauge transformations
of eqs.(12--13) for quark and gluon fields. Hence, the
transformations (14--17) manifest a gauge symmetry of the
theory, and the $R_a$ are the generators of the
transverse residual gauge transformations eqs.(12--13).

{\em 3. Residual gauge invariance and gauge fixing}.
However, the above derivation is not consistent with
the choice of antisymmetric boundary conditions. From
eq.(21), we see that the residual gauge transformations
generated by $R_a$  break the antisymmetric boundary
condition for $A_a^i$ at longitudinal infinity, and
therefore are not allowed. By using the
antisymmetric boundary condition, the first term
in eq.(9) can be integrated out explicitly and the $R_a$ are
reduced to
\BE
	R'_a = \pm 4 \partial^i A_a^i |_{x^-=\pm \infty}
		- \frac{g}{2} \int_{-\infty}^{\infty} dx^-
		(f^{abc} A_b^i \partial^+ A_c^i + 2
		\psi_+^{\dagger} T^a \psi_+)  .
\EE
We can explicitly show that for the $R'_a$,
\be
	& & [R'_a(x_{\bot}), \psi_+(y^-,y_{\bot})]_{x^+=y^+} =
		g \delta^2(x_{\bot} - y_{\bot}) T^a \psi_+(y^-,
		y_{\bot}), \\
	& & [R'_a(x_{\bot}), A_b^i(y^-,y_{\bot})]_{x^+=y^+} = -i g f^{abc}
		\delta^2(x_{\bot} - y_{\bot}) A_c^i (y^-,y_{\bot})
		\nonumber \\
	& & ~~~~~~~~~~~~~~~~~~~~~~~~~~~~~~~~~~~~~~~~~~~~
		+ i \eta(y^-) \delta_{ab} \partial_x^i \delta^2 (x_{\bot}
		- y_{\bot}), \\
	& & [R'_a(x_{\bot}), R'_b(y_{\bot})]_{x^+=y^+} = i g f^{abc}
		\delta^2 (x_{\bot} - y_{\bot}) R'_c(x_{\bot}).
\ee
The commutator of eq.(25) is different from eq.(21) by a
factor of $\eta(y^-)$ in the second term.  From the definition of
$\eta(y^-)$ [see eq.(8)], it follows that the residual gauge
transformations generated by $R'_a$ preserve the antisymmetric
boundary condition.  But, unfortunately, we find that
\BE
	[R'_a ~,~ H] \neq 0 .
\EE

There are two possible interpretations for eq.(27): either eq.(27)
indicates that the residual gauge invariance is broken due
to the antisymmetric boundary condition of $A_a^i$ at longitudinal
infinity, or it implies that the $R_a'$ are not proper generators of the
residual gauge transformations and that the antisymmetric boundary
condition of $A_a^i$ at longitudinal infinity fixes completely the
residual gauge freedom.

In perturbative light-front QCD, the breaking of residual gauge
invariance is associated with the non-cancellation of light-front
infrared divergences in gauge invariant sectors.
The antisymmetric boundary conditions lead to a principal value
prescription, which regularizes light-front infrared singularities.
At tree level, one can check that the principal value prescription
derived from eq.(3) removes all light-front
infrared singularities.  In loop calculations, with
the principal value prescription, it is known that there still
exist spurious poles, which lead to a mixing of ultraviolet and
infrared divergences.  In these cases, it has been demonstrated that the most
severe divergences (mixing of logarithmic ultraviolet and infrared
divergences) are cancelled in the higher-order corrections to the
scale evolution of the hadronic structure functions
{\cite{Furmanski}}.{\footnote{The light-front spurious
poles can also be removed by use of the Mandelstam-Leibbrandt (ML)
prescription {\cite{MLp}}.  Unfortunately, the ML prescription
cannot be applied directly to light-front quantization in
the Hamiltonian formalism.}}  Therefore, the antisymmetric
boundary condition does not obviously break the residual gauge
invariance.

In fact, the second term in the commutator of eq.(25) indicates
that the residual gauge transformations generated by $R_a'$ are
$x^-$ dependent, which is not allowed, as we have pointed out in
the beginning.  Thus, Eq.(27) strongly suggests that one does not
have additional gauge freedom to choose other $A_a^i$ such that
the resulting Hamiltonian remains invariant.  In other words,
the antisymmetric boundary condition of $A_a^i$ at longitudinal
infinity {\em does} fix the residual gauge freedom in $A_a^+=0$ gauge.
We now consider the nonperturbative consequences of the asymptotic
gauge fields induced by the residual gauge fixing.

For physical states, the energy density must be finite.
{}From eq.(1), we see that this requires at least that the field
strengths $E_a^-$ vanish at light-front spatial infinity.  As
a result, we have the following condition [determined by eq.(3)],
\BE
	\partial^i A_a^i |_{x^-=\pm \infty}
		= \mp \frac{g}{4} \int_{-\infty}^{\infty} dx^-
		(f^{abc} A_b^i \partial^+ A_c^i + 2
		\psi_+^{\dagger} T^a \psi_+) .
\EE
To accommodate eq.(28), the commutation relations of eqs.(5--7) have to be
modified.  Thus, eqs.(24-26) are no longer true and $R_a$
are no longer the generators of the residual gauge transformation.
Furthermore, eq.(28) shows that the $A_a^i$ must satisfy antisymmetric
boundary condition at longitudinal infinity.  Thus, in the
space of physical states, the antisymmetric boundary
conditions completely fix the residual gauge freedom, and the
requirement of finite energy density provides an additional
condition to determine explicitly the non-vanishing asymptotic
physical gauge fields at longitudinal infinity.

It may be worth mentioning here that Chodos used a condition similar
to eq.(28) to try to fix the residual gauge freedom in axial
gauge{\cite{Axialg}}.  However, in axial gauge, the first term
in $R_a$ is proportional to the dynamical variables $E_a^i$, and so
cannot be integrated out.  Therefore the operator identity is very
difficult to solve and the final formulation is too complicated
to be practically useful, as pointed out by Chodos himself.
In the light-front gauge, $E_a^i = -\frac{1}{2}\partial^iA_a^i$,
which are not dynamical variables and their vanishing at longitudinal
infinity is reduced to a constraint for $A_a^i|_{x^==\pm \infty}$
in physical states.

{\em 4. Nontrivial topological property}.  Finally, we shall
show an important consequence of the residual gauge fixing by
using antisymmetric boundary conditions in light-front QCD.
It has been pointed out that if
one chooses the $A_a^+=0$ gauge with symmetric boundary conditions,
the nontrivial structure associated with a topological winding number
cannot be addressed {\cite{Franke81}}.  However, with the symmetric
boundary condition, it is not clear how to fix the residual gauge
freedom, which is associated with the nontrivial structure in light-front
QCD. With the antisymmetric boundary condition, the residual gauge freedom
is fixed, and the topological winding number is determined by the
non-vanishing $A_a^i |_{x^-= \infty} = - A_a^i |_{x^-= \infty}$.

Explicitly, we consider the axial current equation (for zero
quark mass)
\BE
	\partial_{\mu} j_5^{\mu} = N_f \frac{g^2}{8\pi^2}
		\tr (F_{\mu \nu} \widetilde{F}^{\mu \nu}),
\EE
where the axial current is $j_5^{\mu} = \bar{\psi} \gamma^{\mu}
\gamma_5 \psi$, and the dual field strength is $\widetilde{F}^{\mu
\nu} = \frac{1}{2} \epsilon^{\mu \nu \sigma \rho}F_{\sigma \rho}$.
The winding number in LFQCD is defined as the net charge between
$x^+= - \infty$ and $x^+ = \infty$,
\BE
	\Delta Q_5 = N_f \frac{g^2}{8 \pi^2} \int_M d^4x \tr
		(F_{\mu \nu} \widetilde{F}^{\mu \nu}).
\EE
The integration on the r.h.s.~of the above equation is defined in
Minkowski space ($M$) and can be replaced by a surface integral.  It can
be found {\cite{zhang2}} that
\BE
	\Delta Q_5 = -N_f \frac{g^2}{ \pi^2} \int dx^+ d^2x_{\bot}
		\tr \left. (A^- [A^1, A^2]) \right|_{x^-=-\infty}^{x^-
		=\infty} ,
\EE
where $A_a^- |_{x=\pm \infty}$ is determined
by eq.(3) and satisfies the antisymmetric boundary condition.
Eq.(31) shows that a non-vanishing $\Delta Q_5$ is generated
from the asymptotic fields of $A_a^i, A_a^-$ and their antisymmetric
boundary conditions at longitudinal infinity.

For physical states, it is particularly interesting to see from
eq.(28) that the asymptotic physical gauge fields are
generated by the color charge densities integrated over $x^-$.
Thus, the topological winding number in $A_a^+=0$ can be explicitly
explored from eq.(28).

To summarize, as a consequence of residual gauge fixing
by the antisymmetric boundary condition, the light-front QCD
vacuum is trivial. Nontrivial QCD features for physical
states are switched
to the field operators and are manifested in the asymptotic
behavior of physical gauge fields at longitudinal infinity.
The trivial vacuum with nontrivial field variables
in the light-front QCD may provide a practical
framework for describing hadrons. A detailed discussion
will be published separately {\cite{zhang2}}.

\section*{Acknowledgement}
We would like to acknowledge fruitful discussions with R. J.
Furnstahl, R. Mills, R. J. Perry, J. Shigemitsu, and K. G. Wilson.
We would also like to thank R. J. Furnstahl and H. Georgi for
their comments and suggestions for improving the manuscript.
This work was supported by National Science Foundation of United
States under Grants No. PHY-9102922, PHY-8858250 and PHY-9203145.


\end{document}